# Plasmonic Particle Integration into Near-Infrared Photodetectors and Photoactivated Gas Sensors: Towards Sustainable Next-Generation Ubiquitous Sensing

*Hendrik Schlicke\*, Roman Maletz, Christina Dornack, Andreas Fery\**

Dr. H. Schlicke
Leibniz Institute for Polymer Research Dresden, Hohe Straße 6, 01069 Dresden, Germany.
E-mail: schlicke@ipfdd.de

Dr. R. Maletz, Prof. Dr. Ch. Dornack
TUD Dresden University of Technology, Faculty of Environmental Sciences, Institute of Waste Management and Circular Economy, Pratzschwitzer Straße 15, 01796 Pirna, Germany

Prof. Dr. A. Fery
Leibniz Institute for Polymer Research Dresden, Hohe Straße 6, 01069 Germany.
TUD Dresden University of Technology, Physical Chemistry of Polymeric Materials, Bergstraße 66, 01069 Dresden, Germany.
E-mail: fery@ipfdd.de

Current challenges in environmental science, medicine, food chemistry as well as the emerging use of artificial intelligence for solving problems in these fields require distributed, local sensing. Such ubiquitous sensing requires components with (1) high sensitivity, (2) power efficiency, (3) miniaturizability and (4) the ability to directly interface with electronic circuitry, i.e., electronic readout of sensing signals.

Over the recent years, several nanoparticle-based approaches have found their way into this field and have demonstrated high performance. However, challenges remain, such as the toxicity of many of today's narrow bandgap semiconductors for NIR detection and the high energy consumption as well as low selectivity of state-of-the-art commercialized gas sensors. With their unique light-matter interaction and ink-based fabrication schemes, plasmonic nanostructures provide potential technological solutions to these challenges, leading also to better environmental performance.

In this perspective we discuss recent approaches of using plasmonic nanoparticles for the fabrication of NIR photodetectors and light-activated, energy-efficient gas sensing devices. In addition, we point out new strategies implying computational approaches for miniaturizable spectrometers, exploiting the wide spectral tunability of plasmonic nanocomposites, and for selective gas sensors, utilizing dynamic light activation. The benefits of colloidal approaches for device fabrication are discussed with regard to technological advantages and environmental aspects, which have been barely considered so far.

# 1. Introduction

Current challenges in environmental science (climate change, biodiversity crisis, environmental pollution/air quality) and medicine (aging society, pandemic threats), as well as the emerging use of artificial intelligence for solving problems in these fields require distributed, local sensing. Thereby, a strong demand for miniaturizable chemical analysis means has grown, e.g., for realization of portable devices for point-of-care diagnostics or the distribution of low power[1] sensors for ubiquitous sensing, via spectrochemical analysis or detection of gaseous species, such as $O_2$, $O_3$, CO, $CO_2$, $H_2O$, $H_2S$, $NH_3$, $NO_x$, $SO_2$, $CH_4$, and the diverse volatile organic compounds (VOCs).[2] Components that are sought to be employed in these fields require (1) high sensitivity, (2) power efficiency, (3) miniaturizability and (4) the ability to directly interface with electronic circuitry, i.e., electronic readout of sensing signals.

Over the last 20 years, particle-based materials have shown an enormous potential for facilitating chemical and bio sensing modalities. Here, much attention has been devoted to optical sensing approaches,[3] e.g., decoration of plasmonic particles with metal-organic frameworks, which leads to a concentration of gaseous analytes in proximity to the particle surface, rendering them effective for detection of low analyte concentrations.[4,5] Presumably most prominently, particle-based materials were utilized for powerful surface-enhanced-Raman based sensing (SERS) including metasurface approaches.[6–9] These can also be extended towards high-sensitivity enantioselective detection of biomolecules.[10]

While SERS and other optical sensing techniques allow for detailed chemical analysis, e.g., via vibrational spectroscopy, they require complex, bulky and expensive spectrometers for readout. In this perspective article we focus on new sensing approaches exploiting the integration of plasmonic nanoparticles into new device types having a strong potential to fulfill the above requirements. Within this context, we also discuss sustainability aspects of functional, particle-based materials for device fabrication.

For miniaturizable chemical analysis, on the one hand, optoelectronic devices, such as portable spectrometers[11,12] are versatile tools. Attracting growing interest, especially the near-infrared (NIR) spectral range may significantly advance possibilities for quick and non-destructive analysis of materials or for noninvasive diagnostics. With wavelengths ranging between 0.8 - 2.5 µm this range encompasses vibrational overtones and combination bands that are characteristic for organic compounds and can well be used for identification and quality assessment of products.[12,13] For the recycling industry, separation processes using NIR-detection are the key technology generating secondary material streams for a circular economy, avoiding resource depletion and other environmental burdens.[14]

Significant advances have been made in introducing semiconductor nanoparticles in various types of optoelectronic devices and sensors related to NIR technologies. Over the recent years multiple research groups have successfully reported the fabrication of highly performant quantum dot based photodetectors,[15,16] culminating in the demonstration of pixel arrays for spatially resolved NIR detection.[17,18] In display industry current research efforts address the fabrication of electronically driven quantum dot light emitting diodes, which are very promising candidates for the successor to commercialized OLED technology. To this end, introduction of nanocrystalline materials with bandgap energies in the NIR spectral range[19] enable light sources that could be well suited for chemical analysis. However, typical narrow-bandgap semiconductor materials are often exotic/costly and frequently contain toxic compounds such as Pb, Hg, Cd and As, which are harmful to human beings and the environment. In part, these materials have been strongly regulated recently by governmental policies, such as the European RoHS regulation.[20] Environmental aspects have to be considered more and more in developing electronic applications.[21] Plasmonic sensors address this, having a reduced material footprint and lower human and ecotoxicity potential according to environmental impact categories, compared to toxic semiconductor materials.[22]

On the other hand, current research aims at the fabrication of low-power but sensitive and selective chemical sensors, which enable the analysis of gas/VOC mixtures. While classical gas analysis relies on stationary devices, such as gas chromatographs and mass spectrometers, portable devices may enable point-of-care diagnostics, e.g., via monitoring a patient's breath for identification of potential pathologic conditions[23,24] or the formation of ubiquitous sensing networks for environmental monitoring. Further, such devices could allow for cheap and easily available monitoring of VOC mixtures emitted by foodstuff upon ripening or degradation, e.g., during harvesting, transport and storage. Studies suggest that the combination of VOC analysis and above described NIR spectroscopy is very promising for food quality assessment.[13] Today's commercialized gas sensors commonly rely on chemiresistive metal oxide (MOX) structures, which have to be operated at high temperatures.[25,26] While showing excellent sensitivity, the necessity of heating the MOXs is a significant drawback when it comes to energy consumption, especially in portable applications or ubiquitous sensing.[2] Even though technology has decreased the power needs of semiconductor gas sensors, current devices still require several mW during operation.[27] As an alternative to the introduction of energy via Joule heating, photoexcitation,[26] e.g., using micro-light plates, has been demonstrated to be a valuable alternative, able to reduce power consumption of gas sensors to or even below the µW range.[27,28] However, typical MOX materials commonly have large band gap energies,[29] requiring UV light sources.

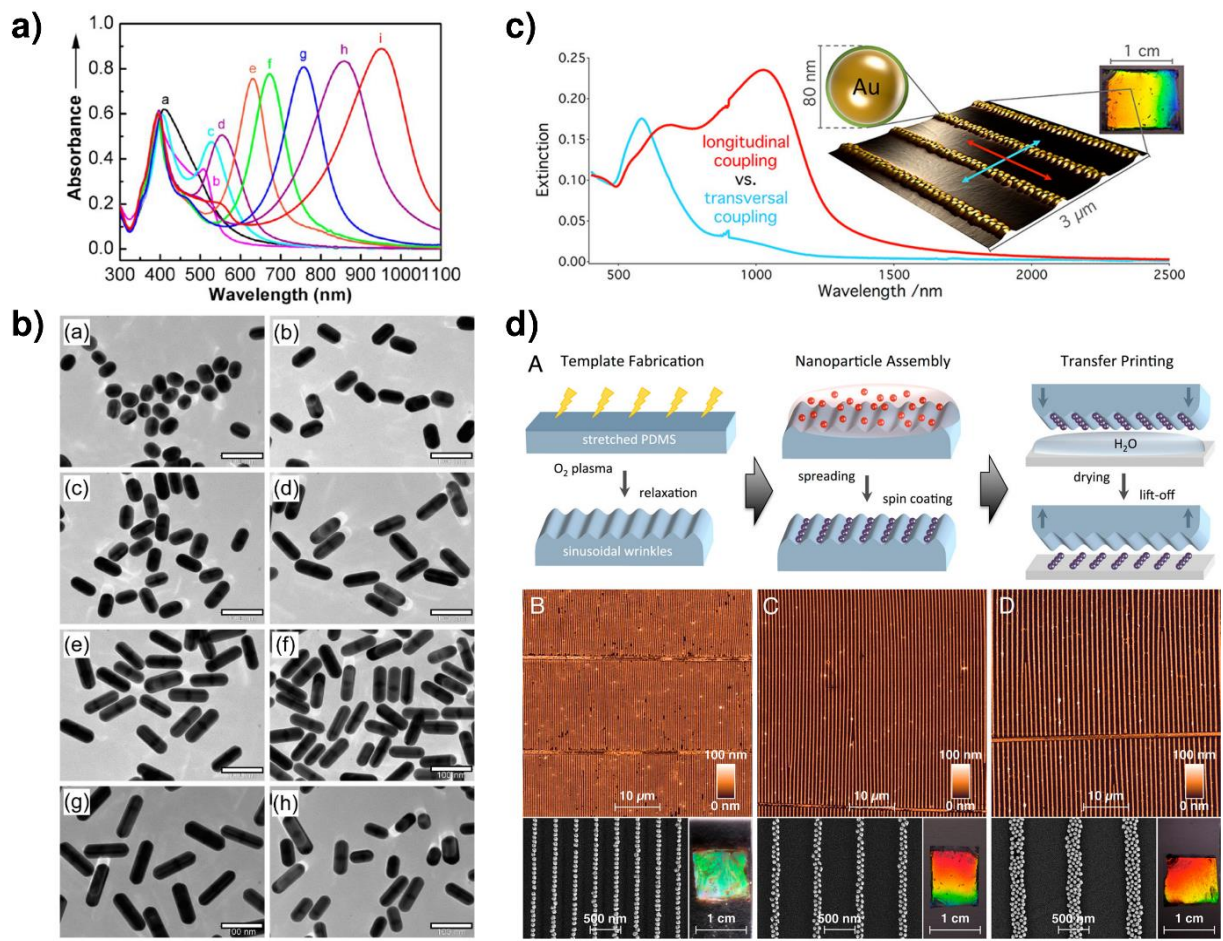

**Figure 1.** Tunability of the absorbance spectra of plasmonic nanoparticles and their assemblies. a) Adjusting the particle composition and geometry: Absorbance spectra of Au@Ag core-shell nanorods (top) with varying aspect ratios from 1.2 (a) to 4.3 (i) and b) corresponding transmission electron microscopy images (bottom). Reprinted with permission from Li et al.[30] Copyright 2013 American Chemical Society. c) Extinction spectra of line assemblies of plasmonic particles, showing longitudinal and transversal coupling effects. d) Fabrication of such particle assemblies through templated assembly. Reprinted from Hanske et al.[31]

## 2. Plasmonic Nanostructures

Both applications discussed above can significantly benefit from introducing plasmonic nanoparticles into upcoming device generations. Plasmonic nanoparticles, most prominently known from noble metals such as gold or silver, may be obtained via colloidal synthesis in high definition.[32] By adjusting their shape (e.g. dot/rod/cube), size and composition, widely tunable absorbance characteristics can be achieved as exemplarily shown for Ag@Au nanorods with different aspect ratios (Figure 1a).[30]

Being the origin for their interaction with light, localized surface plasmon resonance (LSPR) refers to the coupling of coherent oscillations of free electrons to incident light. LSPR is a highly efficient form of light-matter interaction. E.g., in the case of gold or silver nanoparticles, extinction cross-sections significantly exceeding their physical dimensions[33] are observed and enable focusing photon energy to nanoscopic sites.

Besides making use of the LSPR expressed by individual nanoparticles, their assembly,[31,34] e.g., via template-assisted or self-assembly driven methods, may lead to collective inter-particle coupling effects that result in further tunable and narrow spectral features.[31,34,35] An exemplary line pattern of plasmonic nanoparticles obtained via template-assisted deposition and its resulting absorbance characteristics are displayed in Figure 1d and c, respectively.[31]

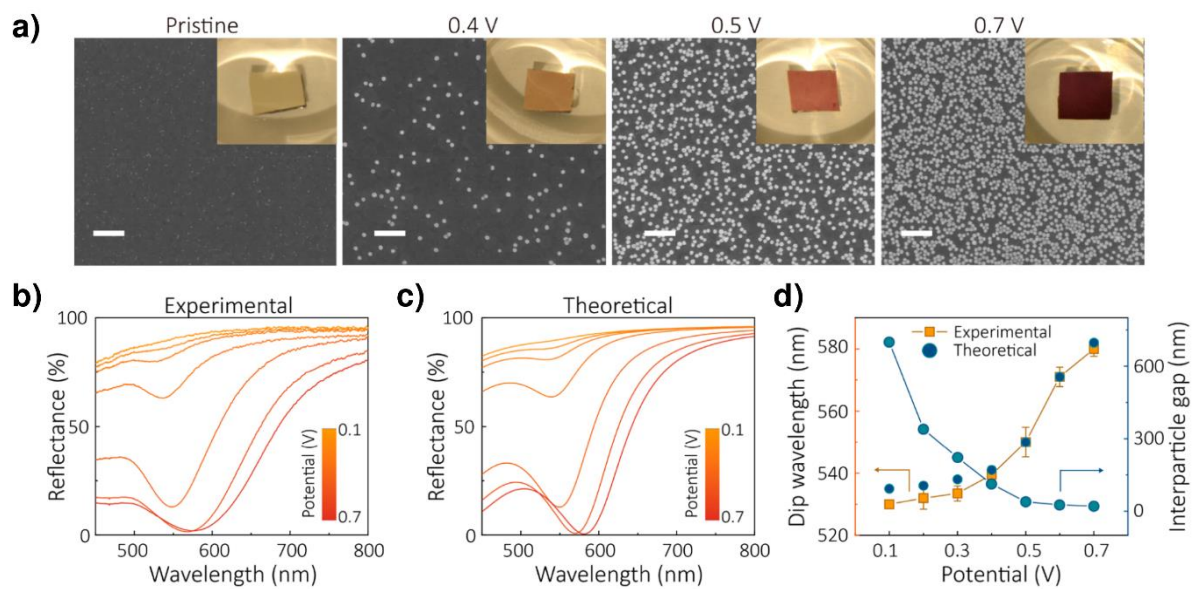

**Figure 2.** Assembly of 40 nm sized, spherical gold nanoparticles on a Ag/TiN electrode, tunable via setting the interfacial potential. a) Ex-situ scanning electron microscopy images of assemblies obtained by applying different interfacial potentials. Scale bar: 200 nm. b) Experimental and c) theoretically fitted reflectance spectra of the nanoparticle assemblies, depending on the interfacial potential. d) Plots of the experimentally determined, potential dependent spectral dip feature wavelength and corresponding computed values for inter-particle gap widths. Reprinted with permission from Ma et al.[36] Copyright 2020 American Chemical Society.

Moreover, different research groups have reported electrochemical strategies capable of reversibly steering particle assembly to achieve electrovariable optical metamaterials.[37] For example Ma et al. impressively demonstrated the controlled assembly and disassembly of 40 nm sized gold nanoparticles on a Ag/TiN electrode in an electrochemical cell.[36] Figure 2a depicts (ex-situ) scanning electron microscopy (SEM) images showing the electrode surface covered with nanoparticles upon positively polarizing it using a varying externally applied potential. Hereby the surface coverage and inter-particle spacing could be reversibly tuned, resulting in a significantly altered reflectance behavior (Figure 2b), which was in good agreement with theory (cf. Figure 2c,d). Markedly, the reflectance of the electrode surface could be almost entirely quenched at the plasmonic resonance wavelength of the particle assembly. Electronically tunable plasmonic metasurfaces enable a (even non-linear) variation of their spectral properties, which gives rise to manifold applications as tunable optical components, such as filters, or as multispectral components integrated in plasmonically enhanced optoelectronics.

After excitation of LSPR, three effects, which also have been intensely discussed regarding photocatalysis recently,[38–40] can be utilized to make use of the collected photoenergy in devices: (1) Strong enhancement of electric near fields in proximity of the plasmonic particles, through the di- and multipolar oscillations of conduction band electrons. Such near fields can be exploited to excite proximal species, e.g. through plasmonic resonance energy transfer (PRET).[41,42] (2) Generation of hot carriers/electrons.[40] These are charge carriers with kinetic energies of up to 1-4 eV,[43] formed in the plasmonic particles after excitation, and are capable of overcoming potential barriers (e.g. Schottky junctions). This enables their injection into peripheral species, such as semiconductors. (3) Photothermal heating due to lattice vibrations induced by electron-phonon coupling.[44] Hereby, locally, high temperatures can be achieved. Being a prominent research focus for photothermal therapy,[45] the generated heat can also be used for transduction in novel sensors.

Device integration of plasmonic nanostructures with semiconducting nanomaterials or polymers is a promising approach for creating highly functional and environmentally friendly devices for chemical analysis (Figure 3). In the field of NIR spectroscopy, integration of plasmonic nanoparticles opens great opportunities for extending and tuning spectral characteristics, as well as for multispectral detection and imaging, which, in combination with spectral reconstruction and machine learning, enables computer-aided spectroscopy. In the field of chemical gas sensing, integration of plasmonic particles is expected to enable light-driven devices, which are energy-efficient and therefore could enable the assembly of ubiquitous sensing networks and portable devices. Additionally, the application of modulated light sources and evaluation of the time-dependent photoactivated sensor responses is a promising method for analyte recognition via machine learning/artificial intelligence approaches.

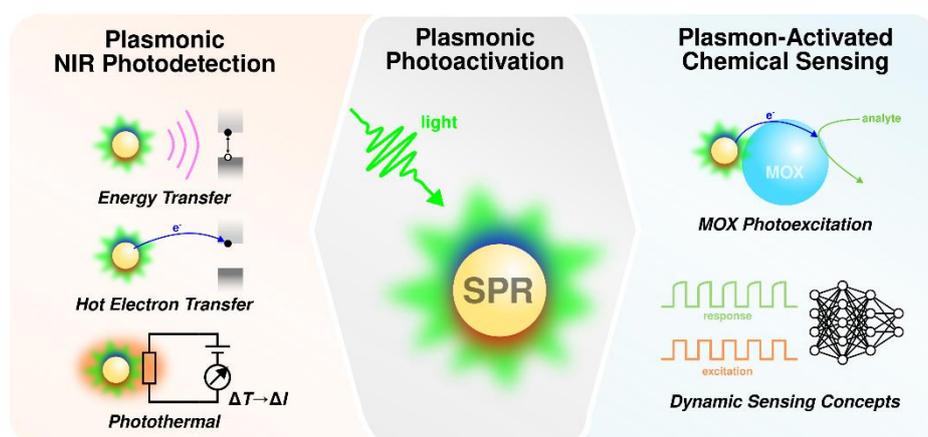

**Figure 3.** Photoactivation through plasmonic nanoparticles offers multiple opportunities for next-generation devices in the fields of NIR photodetection and chemical gas sensing.

Within this perspective, we provide an overview of recent developments in utilizing plasmonic effects in such novel nanoparticle-based devices. While more technically detailed discussions may be found in recent review articles on plasmonic photodetection[46] and photoactivated gas sensors,[29,47] we aim to highlight recent studies pointing out applications of plasmonic particles in sensors, especially suitable for chemical analysis. In the first part, we discuss recent approaches aiming at the use of plasmonic particles for NIR photodetection and multispectral sensing, while in the second part recent research works on optically activated conductometric gas sensors will be addressed.

## 3. Plasmon-Enabled Devices for NIR Photodetection and Multispectral Sensing

Today, silicon is the first choice when it comes to commercial photodetectors for the visible (and proximal NIR) wavelength range. Over the past years elaborate and highly controlled fabrication technologies for structuring silicon with critical dimensions down to the low nm scale[48] have been established, peaking in the fabrication of processors with billions of highly-performant transistors on several-inch scale wafers. Accordingly, silicon photodiodes can easily be fabricated following standard CMOS processes[49] and are abundant, extremely cheap, down-scalable to a few microns for high-resolution photodetector pixel arrays, and highly effective. The operational range of silicon-based photodetectors is however limited due to its intrinsic band gap of $E_g \sim 1.1$ eV.[50] Photons with energies in the NIR range below Si's band gap energy (>1100 nm) are incapable of exciting charge carriers within the bulk Si material, leading to photocurrent. Unfortunately, as described above, the detection limit of Si photodetectors marks the onset of the NIR spectral range, that is currently drawing significant interest as it contains valuable spectral features for qualitative and quantitative assessment of substances and products. To address this wavelength range with conventional photodetectors operated in photoconductive or photovoltaic mode, narrow band-gap semiconductors are required. While current approaches target the development of new material systems, common materials used in commercialized devices and in recent works on quantum dot based systems encompass, e.g., Ge, InGaAs, InAs, PbS, PbSe, InSb, HgCdTe, HgTe.[51] Many of these narrow-band gap NIR semiconductor materials are either expensive or contain toxic elements, which have undergone strict regulations recently. Hence, there is a strong need for inexpensive and environmentally friendly solutions extending the silicon wavelength range of photodetectors to the NIR. Replacement can lead to reduced environmental impact of those functional materials.[52]

With the unique light-matter interaction of plasmonic structures the nanoscale world offers another, highly efficient option to collect photoenergy, as an alternative to the direct excitation of semiconductor structures. To make use of the collected energy, it has to be transformed into an electronic signal that could be recorded and interpreted by readout electronics. In the following, we highlight three strategies that utilize plasmonic effects therefore, involving (1) the near field enhancement of plasmonic particles and energy transfer to proximal (semiconductor) species leading to photocurrent, (2) the transfer of hot carriers to peripheral species inducing photocurrent and (3) detection based on heating effects, exploiting plasmonic photothermal processes.

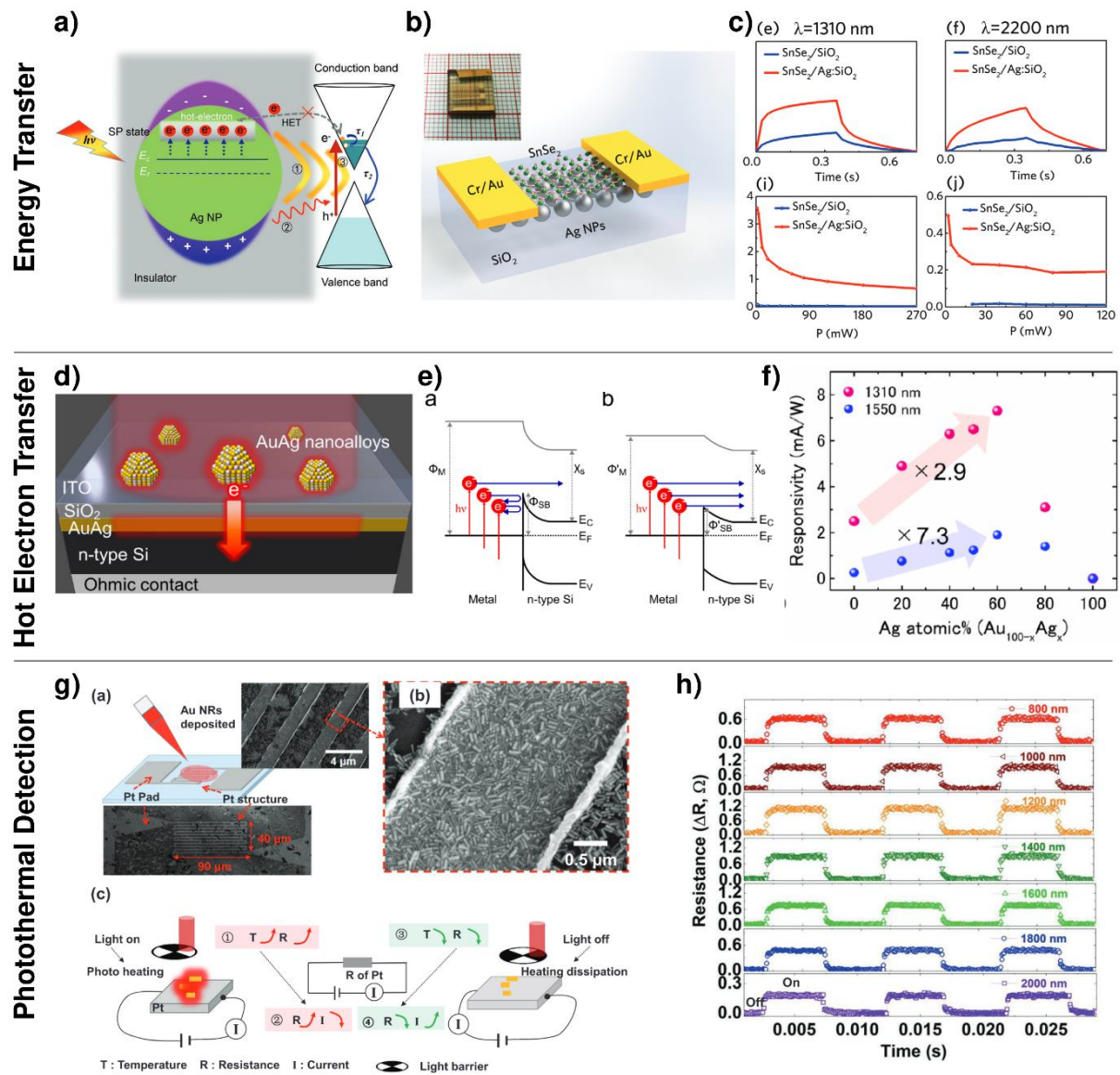

**Figure 4.** Different approaches for utilizing plasmonic nanoparticles and nanoparticle assemblies for NIR photodetection. a) Sensitization of semiconductor/2D materials through plasmonic local field enhancement and energy transfer. b) Schematic of a SnSe$_2$ photodetector with proximal plasmonic Ag nanoparticles. c) Photoresponse and sensitivity of SnSe$_2$/AgNP:SiO$_2$ devices to illumination with NIR light. Reprinted from Sun et al..[53] Copyright 2021 Wiley‐VCH GmbH. d) Schematic of plasmon-induced hot-carrier transfer into semiconductor materials. e) Schottky junction formed at a metal/silicon interface. The barrier height was tuned by adjusting the alloy composition of the plasmonic nanoparticles. f) Enhancement of photodetector NIR responsivity achieved by tuning the Schottky barrier height via alloying of the plasmonic material. Reprinted with permission from Okamoto et al.[54] Copyright 2024 American Chemical Society. g) Photothermal plasmonic photodetector based

on a broadband-absorbing assembly of plasmonic nanorods deposited onto a platinum thermistor and h) its resistance response to exposure with NIR light of varying wavelength. Reprinted from Xiang et al.[55] Copyright 2021 Wiley‐VCH GmbH.

*Near-field enhancement and energy transfer.* Recently, numerous reports of (opto-)electronic components from graphene and later 2D materials were published. The studies highlighted the manifold advantages of these new materials over bulk semiconductors, e.g., extraordinary electronic properties[56] and flexibility[57]. When applied for light detection, however, they show insufficient absorption due to their low thickness, (few- to monolayers), hindering the fabrication of efficient photodetectors. Due to their large extinction cross-sections, plasmonic particles[53,58–60] and their assemblies may serve as concentrators for incident light, and have shown tremendous enhancements in photodetector responsivity.

For example, Luo et al.[60] fabricated high-performance NIR photodetectors based on plasmon-enhanced single-layer graphene / InP Schottky junctions. The intrinsically low NIR absorption of the formed diodes was improved by the introduction of $SiO_2$ coated Au nanorods with a pronounced longitudinal plasmon absorbance band in the NIR range. While the $SiO_2$ shell prevents hot electron transfer from the plasmonic particles to the semiconductor structure, the authors attribute the observed strong enhancement of responsivity to efficient light trapping in the nanorods in proximity to the semiconductor and increased carrier generation due to the strong local field enhancement.[60]

In another example, Sun et al.[53] reported the fabrication of few-layer $SnSe_2$ photodetectors on $SiO_2$ substrates with Ag nanoparticles embedded right below their surface (cf. Figure 3b). While again not being in close contact with the nanoparticles, impeding hot electron transport, the $SnSe_2$ material showed a strong enhancement of its NIR photoresponse (cf. Figure 3c). This was attributed to plasmonic resonance energy transfer (PRET) and local field enhancements due to the proximal nanoparticles (Figure 3a) by the authors.

These recent works highlight the versatility of NIR photodetection approaches relating to sensitivity improvement by locally enhanced plasmonic near fields and energy transfer processes. The on-chip combination of a universal (e.g., 2D) acceptor material with plasmonic nanoparticles and their assemblies having different absorption characteristics could lead to the fabrication of monolithic, multispectral sensing devices that could facilitate cost-effective and portable NIR spectroscopy.

*Hot electron transfer.* Upon combining semiconductor and metal structures, depending on the metal's work function and the semiconductor's doping state, Schottky junctions might form at their interface. Herein, as exemplarily depicted in Figure 3e, transfer of charge carriers from the semiconductor to the metal may occur, resulting in the formation of a barrier, which eventually hinders charge transport. If the metal side of the semiconductor/metal junction is a plasmonic component, incident light may result in the excitation of LSPR. The high kinetic energy of hot electrons resulting from this process enables them to overcome the Schottky barrier and to be injected into the semiconductor, leading to photocurrent.[61–63] Due to the energetic alignment of the junction, photon energies required for this process may be significantly lower than the semiconductor's band gap, enabling efficient photodetection in the longer wavelength range.

NIR photodetection enabled by hot electron transfer from plasmonic nanoparticles has been demonstrated with different grating structures that are in direct contact with semiconductors[64,65] or using silicon pyramids decorated with gold nanospheres, showing pronounced absorption at energies below Si's band gap.[66] In a recent work, Okamoto et al. demonstrated the fabrication of improved plasmonic nanoparticle-based Schottky-type Si /

Ag$_x$Au$_{1-x}$ nanoparticle photodetectors for the NIR range (Figure 4d), which they obtained by optimizing the work function of plasmonic structures through alloying. Hereby, the height of the Schottky junction was lowered (Figure 4e) and efficient detection of NIR radiation (Figure 4f) was achieved. In their approach the nanoparticles were integrated into their devices via cathodic arc plasma deposition.[54]

*Photothermal transduction.* Another approach for achieving NIR photosensitivity discussed here involves using photothermal effects along with device elements that convert heat to electronically interpretable signals.[67,68] Xiang et al. recently reported the fabrication of a broadband NIR photodetector (see Figure 4g).[55] The authors assembled gold nanorods of different geometries to obtain, through their individual absorbance characteristics as well as due to coupling effects, broadband NIR absorbance. Particle assembly was conducted on top of a microfabricated platinum thermistor, which showed an increase in resistance due to a temperature rise caused by plasmonic photothermal heating when the sensor was exposed to NIR light.

Yang et al. demonstrated that this detection principle could be employed to achieve spectral selectivity by deposition of gold nanorods with differing spectral absorbance characteristics onto thermistors.[68] The versatility of using the photothermal concept was further shown by Mikkelsen and co-workers. In 2020 the group reported the fabrication of photothermal detectors by depositing a metasurface consisting of a metal layer, decorated with silver nanocubes, onto a pyroelectric material. By adjusting the cube geometries, widely tunable absorption characteristics in the visible and NIR regions were obtained. In contrast to other, commonly slow photothermal photodetectors, their devices showed impressively short response times in the ns range and bandwidths up to 300 MHz, which are approaching values of semiconductor-based devices.[69] Responsivities of the sensors were as high as ~170 mV/W and 0.33 mV/V for light wavelength of ~900 nm and 1900 nm, respectively. To estimate the thermal-temporal limits of the detectors, the authors combined electromagnetic simulations with solid-state heat transfer simulations. Hereby, they estimated temperature changes experienced by the pyroelectric material of multiple K within the picosecond range in their devices. They conclude that the observed response times are hence not limited by the thermal transfer process, but rather due to the electronic time constant arising from their present device architecture.[69]

The above works clearly demonstrate that we can effectively tune near-infrared (NIR) absorbance characteristics by using tailored colloidal nanoparticles and their assemblies. These structures, when properly integrated into devices, have the potential to create highly adaptable next-generation photodetectors for the NIR range.

The universality of readout methods with respect to their combination with plasmonic structures – as shown in the case of photothermal readout – calls for the fabrication of multi-pixel detector arrangements, i.e., arrays of multiple, individually addressable pixels that are functionalized with plasmonic materials having varying spectral characteristics. By adjusting the spectral properties of individual pixels to match characteristic absorbance features of target analytes or marker molecules, selective detection or even spatial resolution/imaging can be realized. The latter requires fabricating an array of pixel groups consisting of multispectral subpixels.

When a multitude of different pixels with semi-continuously varying narrow-band spectral properties are fabricated, incoming light can be resolved as a spectrum (cf. Figure 5a, center). Still, narrowband detection is not necessarily required for this purpose: If pixels with arbitrary, even broadband, but differing responsivities are supplied, computer-aided spectroscopy, i.e., making use of spectral reconstruction algorithms[70] or machine learning approaches may enable monolithic on-chip spectrometers, operable without any moving components (cf. Figure 5a, bottom).

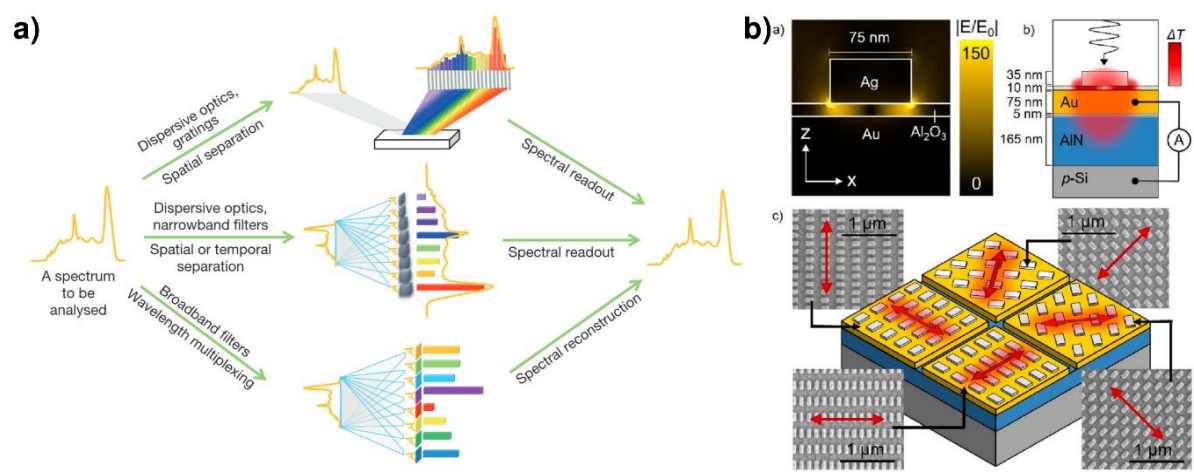

**Figure 5.** a) Concepts for the realization of spectrometers using dispersive gratings for spatial separations of the light spectrum, multiple narrowband filters (or detectors with narrowband sensitivity) and multiple broadband filters (or pixel detectors with broadband sensitivity). Bao et al.,[70] reproduced with permission from SNCSC. b) Photothermal, polarization-sensitive photodetector based on differently oriented plasmonic metasurfaces deposited onto a pyroelectric material. Reprinted with permission from Wilson et al..[71] Copyright 2023 American Chemical Society.

In addition to spectral resolution, the use of the polarization dependence of plasmonic structures enables the fabrication of polarization sensitive detectors: Such approaches were demonstrated by Wilson et al.[71] Using the device structure described above for pyroelectric photothermal sensing, they prepared four detector pixels with plasmonic metasurfaces, consisting of ordered arrays of anisotropic silver nanostructures having different orientations (Figure 5b). Utilizing the polarization sensitivity of the absorber materials, the group achieved a plasmonic photothermal detector that is capable of resolving the polarization of incident light.

## 4. Plasmon-Activated Conductometric Gas Sensors

The working principle of today's commercialized gas sensors is widely based on the resistive readout of metal oxide (MOX) materials, heated to temperatures of several 100 °C.[25,72] Here, ambient oxygen is ionosorbed at the MOX surface, whereas the dominating species ($O_2^-$, $O^-$, $O^{2-}$) typically depends on the operating temperature.[25,72] Due to the surface binding, electrons are withdrawn from the MOX materials, inducing changes in the surface carrier concentrations. This effect then causes - depending on the materials' majority carrier type - a decrease (n-type) or increase (p-type) in conductance of a sensing element. Similar effects occur upon sorption of other oxidizing gases such as $NO_2$ or $O_3$.[25,72] Upon exposure to reducing gases, these react with the ionosorbed oxygen species, leading to reinjection of the bound electrons into the MOX material, resulting in an conductance increase (n-type) or decrease (p-type) of the MOX material, being the sensing signal.

The heating required for achieving the sensors' operation temperature causes the vast amount of power required for operating MOX gas sensors. While recent developments, e.g., heating using free-standing micro-hotplates, enabled a reduction in thermal capacity and hence in energy consumption, power levels in the mW range are still required for operation.[27] As an alternative to Joule heating of the sensing materials' substrates, excitation of MOX sensors with light instead of heat was reported.[27–29,47] However, typical MOXs are wide-bandgap materials[29] and hence require high energy (UV) photons for operation. UV light sources, especially with deep-UV emission, still show low efficiencies,[73,74] are not easy to co-integrate and their high-energy radiation is harmful. Here we would like to highlight a recently emerging approach for shifting the activation wavelength to visible range: The introduction of plasmonic materials.[29,47,73,75–79]

In one early work Xu et al. demonstrated plasmonic light activation of MOX sensors using visible light.[76] The group first fabricated ZnO nanotetrapods via a thermal evaporation process and decorated the structures with gold nanoparticles. Therefore, a 5 nm gold layer was first deposited via physical vapor deposition and subsequently annealed at 500 °C to form nanoparticles on the MOX structures (cf. Figure 6a). The structures showed strongly enhanced responses to exposures to ethanol (inset of Figure 6a), methanol, formaldehyde and acetone under white light (Hg lamp) illumination, already at room temperature.[76]

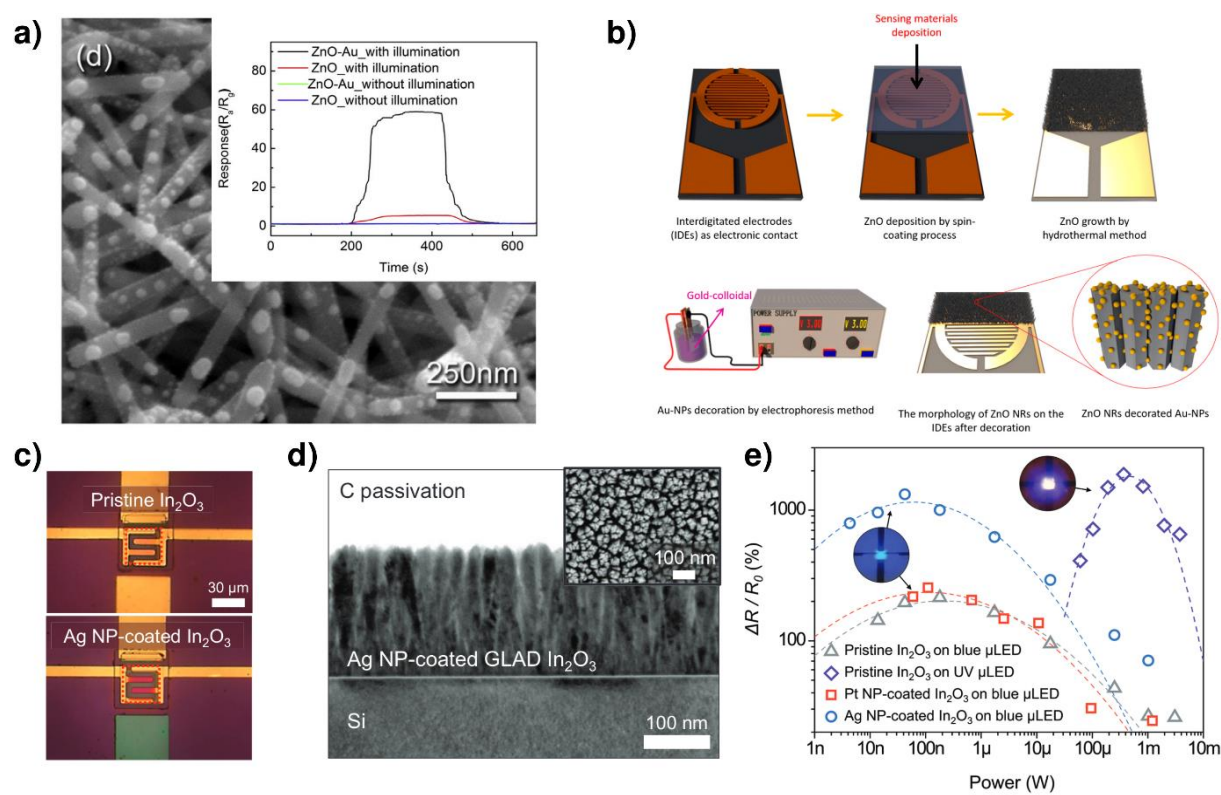

**Figure 6.** a) Scanning electron microscopy images of gold nanoparticle decorated ZnO nanotetrapods. The inset shows the influence of AuNP decoration on the material's chemiresistive response to exposure to 500 ppm ethanol vapor at 17 °C under illumination. Reprinted from Xu et al.[76] Copyright 2018, with permission from Elsevier. b) Fabrication of plasmonically activated gas sensors from ZnO nanorods that were electrophoretically functionalized with colloidal gold nanoparticles. Reprinted from Qomaruddin et al.[80] CC BY 4.0. c) Optical and d) electron microscopy images of $In_2O_3$ based gas sensors, functionalized with Ag nanoparticles via physical vapor deposition, fabricated directly on top of GaN micro-LEDs. e) Response amplitudes of the AgNP-functionalized $In_2O_3$ gas sensors under exposure to $NO_2$ (1 ppm). The AgNPs clearly enabled the use of visible, blue light for sensor activation. Reprinted from Cho et al.[73] Copyright 2023 Wiley-VCH GmbH.

In a different study, Qomaruddin et al.[80] employed electrophoretic deposition to decorate hydrothermally grown ZnO nanostuctures with citrate-stabilized gold nanoparticles that were obtained via colloidal synthesis beforehand (Figure 6b). At room temperature, they observed a 46-fold increase in response (to 891%) of the sensor to $NO_2$ (10 ppm), which they attributed to LSPR activation when operating the chemiresistive structures under visible light excitation. While the field of plasmonically activated gas sensors is still in its infancy, its potential for the fabrication of highly integrated low-energy gas sensors was already impressively demonstrated in a recent paper[73] by Cho et al.. The authors reported the fabrication of sensing devices containing an $In_2O_3$ active layer functionalized with plasmonic Ag nanoparticles, deposited via physical vapor deposition. The chemiresistive $In_2O_3$/AgNP layer was formed directly on top of a GaN μLED to yield a monolithic light-driven sensing device, enabling direct illumination of the sensing material (Figure 6c,d). In contrast to an earlier work by the group,[81] which reported a similar device without plasmonic silver particle decoration and relying on UV-GaN LEDs with emission wavelength below 400 nm for MOX excitation, the introduction of the plasmonic particles enabled the use of more efficient GaN LEDs with a peak emission wavelength at 435 nm. Following this approach the authors highlighted the detection of $NO_2$ (1 ppm) at a low power consumption of only 63 nW (Figure 6e).[73] An overview over material systems used in recent studies for achieving plasmonic activation of MOX gas sensors, employed excitation means and target analytes is provided in Table 1.

While the enhancement of sensor performance through photoexcitation of plasmonic particles incorporated into MOX sensing materials was demonstrated in the above works, its mechanism still requires further investigation. Transfer of photoexcited carriers from the metal to the MOX, which can then take part in surface reactions enabling chemical sensing is commonly discussed.[29,47,73,76,79,80] However, alongside, the introduction of noble metals with MOX gas sensing materials could further contribute to sensor performance via their catalytic properties and spill-over effects.[79,82] Therefore, future research should be further directed to fundamental investigations of effects occurring upon introduction and excitation of plasmonic materials. Here, sensors having highly defined and tunable plasmonic metal/MOX interfaces, obtainable, e.g., via colloidal assembly, are valuable. They should be comprehensively characterized under different excitation means, i.e., thermal excitation vs. monochromatic light excitation at UV wavelength for direct MOX excitation as well as visible excitation with wavelengths centered on and off plasmonic features.

**Table 1.** Material systems, target analytes and excitation light sources used in recent studies on plasmon-activated gas sensors.

| Material system | Target gases (concentrations) | Excitation | Reference |
|---|---|---|---|
| ZnO nanotetrapods / Au nanoparticles | Ethanol (1 ppm-1000 ppm) Acetone (500 ppm) Formaldehyde (500 ppm) Methanol (500 ppm) | Hg lamp | [76] |
| $In_2O_3$ / Ag nanoparticles | $NO_2$ (0.14-5 ppm) | 435 nm integrated GaN µLED | [73] |
| ZnO nanorods / Au nanoparticles | $NO_2$ (10 ppm) | various, 465-640 nm | [80] |
| ZnO nanorods / Au nanoparticles | $NO_x$ (6 ppm) | white light | [75] |
| ZnO /Ag heterostructure nanoparticles | $NO_2$ (0.5-5 ppm) | various, 365-520 nm | [83] |
| $SnO_2$ nanofibers / Au nanoparticles | $NO_2$ (0.125-5ppm) | 450 nm, 530 nm, 630 nm | [79] |
| ZnO Nanorods / Pd nanoparticles | $CH_4$ (0.01-0.1%) | 590 nm (80 °C) | [84] |

Based on these recent, initial studies, it is evident that the introduction of plasmonic particles into gas sensing devices represents a valuable method to efficiently collect and focus photonic energy to nanoscopic sites, enabling overall more efficient sensing schemes. Especially with the rise of µLED and nano-LED technology, miniaturizable light sources become available, which can be co-integrated with chemiresistive sensing materials. Further, for ubiquitous sensing networks, sensor excitation using sunlight might become imaginable, reducing the power requirements of a gas sensing element just to the needs of its readout circuitry.

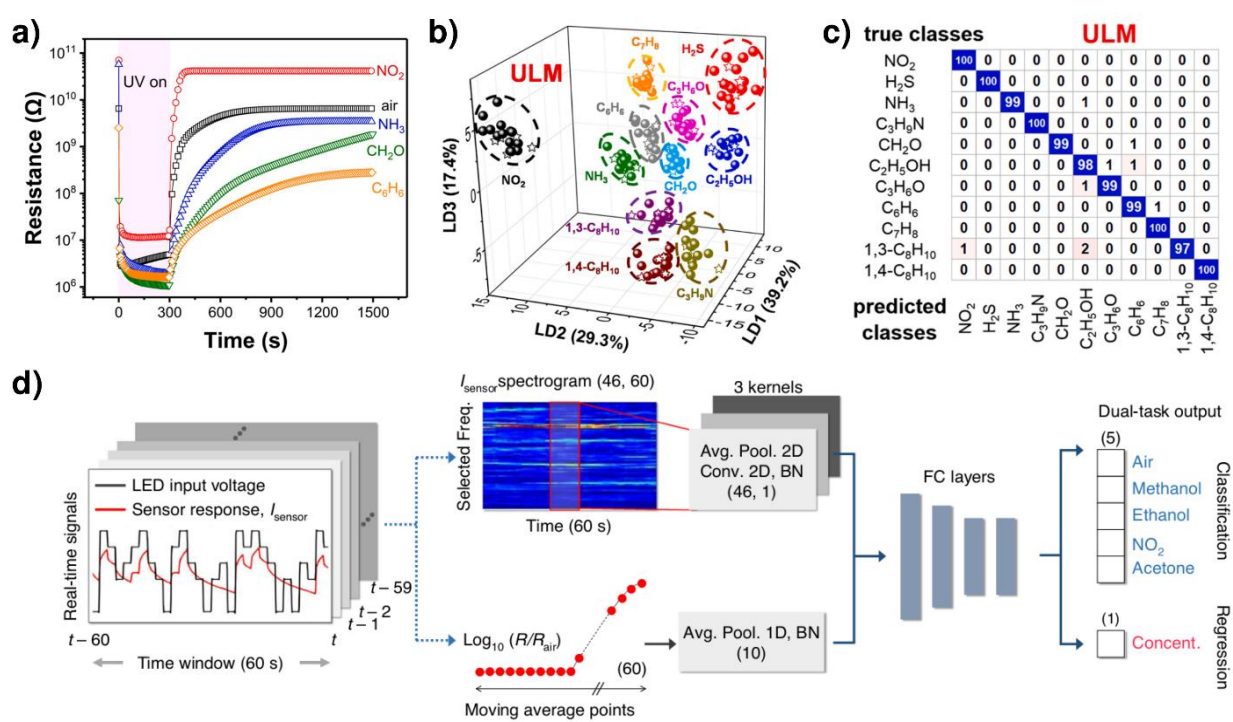

**Figure 7.** Dynamic responses of gas sensors contain valuable information on the present analyte. a) Transient responses of a MOX gas sensor to UV excitation in presence of different analyte species. b,c) Linear discriminant analysis (b) and multilayer preceptor neural network analysis (c) of transient response features of dynamically photoexcited MOX sensors enable analyte recognition. Reprinted from Li et al.[85] Copyright 2022, with permission from Elsevier. d) Utilization of dynamic excitation of an $In_2O_3$/AuNP chemiresistor using a pseudo-randomly pulsed µLED. Machine learning was used for analyte recognition by analyzing frequency components of the dynamic response signal and for concentration determination. Reprinted from Cho et al.[78] CC-BY 4.0.

Aside from reducing power requirements of chemical sensors, light activation gives rise to different options for dynamic sensor operation. Because the light output of common (µ)LEDs can facilely be modulated by adjusting the drive current, even at high frequencies, the transient response of a sensing material to varying excitation intensities may be recorded and correlated. Here, dynamic processes, such as ad-/desorption, mass transport, or chemical reactions,[85–87] that can be on-demand triggered and monitored may contain valuable information on present analytes. Further, phase-sensitive (lock-in) detection schemes could be employed to significantly enhance the signal/noise ratio of the sensor responses.

In an exemplary study, aiming at analyte recognition using room-temperature photoactivated gas sensors, Li et al. reported the use of modulated UV (365 nm) excitation (here without the use of plasmonic materials) of arrays containing $ZnO/(Sr_{0.6}Bi_{0.305})_2Bi_2O_7$ (ZnO/SBO), ZnO and $In_2O_3$ based chemiresistors. They employed multivariate analysis and machine learning for interpretation of the sensors' dynamic response data. Initially, as displayed in Figure 7a, they observed characteristic patterns of the resistance recovery after pulsing MOX sensors with light in presence of gaseous analytes. The authors attributed this behavior to differing analyte molecular properties, i.e., their distinguishable capability to reduce oxygen species photosorbed on the MOX surface beforehand during excitation. Based on these observations the authors employed a rectangular modulation waveform of the excitation signal (continuous exposure with defined off-pulses, correlated with analyte exposure) and recorded the response transients. Following pre-processing the group used principal component analysis, linear discriminant analysis (see Figure 7b) and a multilayer perceptron neural network (cf. Figure 7c) to successfully discriminate between a set of analytes.[85]

The benefits of dynamic photoactivation are also underlined by a recent study of Cho et al. following a slightly different approach. Using only a single µLED driven chemiresistor consisting of $In_2O_3$ functionalized with gold nanoparticles, they reported the identification of four different VOCs and binary (ethanol/methanol) VOC mixtures by operating the excitation LED in pseudorandom modulation and evaluating the sensor response's frequency components using a deep neural network (Figure 7d).[78]

## 5. What's next?

Near infrared spectroscopy and the selective analysis of gases and vapors are key technologies for solving various current and future problems.[12,13,23,24] These encompass the massive waste of food caused by losses, which could be prevented by optimizing harvesting, transport and storage times based on appropriate spectral or chemical information. The early recognition of diseases through suitable diagnostic approaches could enable prompt therapeutic responses, thereby prevent severe courses of disease and reduce treatment costs. Environmental monitoring using ubiquitous sensor networks enabled by energy efficient devices may achieve early detection of forest fires or evolving environmental hazards. Portable, affordable sensing devices could enable end-users to identify potential contamination of foods or the authenticity of products and pharmaceuticals. Above we highlighted a set of recent studies, demonstrating how the introduction of plasmonic materials and the exploitation of their unique light-matter interaction in next-generation NIR optoelectronics and gas sensing devices can significantly improve performance, enable new transduction schemes and fabrication routes without the need for toxic and environmentally harmful materials. Both, the mentioned promising applications and also the considered environmental aspects during development of new functional material solutions, can lead to a strong decrease of environmental burdens like eco- und human toxicological pressure and greenhouse gas relevant emissions.

While different approaches for using plasmonic materials have been fundamentally demonstrated, promotion of device integration strategies leading to higher technology readiness levels, scalability and finally realization of market mature products is now required. A unique advantage in using plasmonic nanostructures is given by the wide spectral tunability, available without changing the material system. In combination with universally compatible transduction schemes, such as the photothermal transduction highlighted in section 3, they should enable multi- and hyperspectral detector components, easily tailorable for specific application scenarios. In the field of gas sensing, introduction of plasmonic nanostructures has shown a great potential for energy-efficient room temperature sensors, dynamically operable with visible light from abundant and inexpensive light sources, or even sunlight.

Particularly taking into account the recent advances of machine learning/AI technologies, optoelectronic devices with highly tunable spectral characteristics or the large feature space accessible by dynamic operation of chemical sensors may, in combination with computationally assisted spectroscopy or advanced recognition algorithms, yield powerful tools.

Most of the initially demonstrated hybrid gas sensor systems rely on plasmonic materials deposited via physical vapor/sputter deposition, lacking high structural definition. In order to achieve better understanding of sensing mechanisms, sensing materials with highly defined (and tunable) geometric features are extremely valuable. Also, for enabling precise spectral tunability in optoelectronic components, high structural definition is required. Here, construction of functional materials via assembling colloidal (e.g. MOX and metallic or hybrid) nanostructures[88] is a worthy tool. Wet chemical fabrication routes provide precise bottom-up design and fabrication of nanostructures as well as cost- and energy-efficient deposition schemes, which also results in environmental benefits. Still, when adhering to colloid-based fabrication procedures, particular care has to be taken regarding the inter-particle interfaces. Colloidal systems are stabilized using surface ligands, which commonly end up being part of a functional layer fabricated from them. Still, besides removal of surface ligands through, e.g., thermal decomposition,[89] enhanced device functionality could be realized by initially introducing functional ligands or polymer matrices with tailored electronic properties.[90]

With the increasing amounts of problematic, partially hazardous e-waste from end-of-use electronic devices, as well as the increasing demand for rare source materials, comprehensive recycling strategies will play an pivotal role in future manufacturing.[91] Extensive consideration of end-of-use scenarios of electronic products is inevitable and strategies for resolving the e-waste problem are required. Here, the integration of elements that allow for controlled deconstruction of devices or device components already during their fabrication will enable recovery of materials and improve reusability and recyclability. Especially rare elements like Au and Ag even when used in smallest amounts have a very high ecological footprint, making the application of reuse concepts very promising.[92]

Besides the above-mentioned technological advantages of colloidal materials and fabrication approaches, the assembly of devices using particle-based materials is expected to result in improved reusability and recycling of device components in the near future. Via tailored ligand and polymer chemistry the selective disassembly of colloidal device components might be triggered by immersion into solvents, controlled changes of temperature, pH, or other, more specific chemically induced cleavage of inter- and intramolecular bonds. The applied triggers can be chosen by their environmental performance using Life Cycle Assessment methodology allowing to reduce environmental impacts of the processes.[93]

Using selective chemistry, orthogonal binding schemes may be introduced and different device components could be detached from the device structures sequentially, allowing for good separation of material classes improving reusability. Especially because the plasmonic devices typically involve rare elements such as noble metals that are currently scraped via elaborate procedures and fabrication of precisely tailored plasmonic nanostructures involves cumbersome syntheses, recuperation of these particle-based materials has a great significance.

**Acknowledgements**

H.S. and A.F. acknowledge financial support from the Deutsche Forschungsgemeinschaft (DFG) within RTG2767, project no. 451785257.